\documentclass[reprint,amsmath,amssymb,aps]{revtex4-2}

\usepackage{graphicx}  
\usepackage{color}
\usepackage{dcolumn}
\usepackage{bm}

\begin{document}

\preprint{AIP/123-QED}

\title{ 
Transport of ultracold dipolar fermions in one-dimensional optical lattices}

\author{Barnali Chakrabarti$^{1,2,3}$, N D Chavda$^4$, Andrea Trombettoni$^{5,6}$, Arnaldo Gammal$^2$}
\affiliation{$^1$ Department of Physics, Presidency University, 86/1 College Street, Kolkata 700073, India.}
\affiliation{$^2$ Instituto de Física, Universidade de S\~ao Paulo, CEP 05508-090, S\~ao Paulo, Brazil.} 
\affiliation{$^3$ The Abdus Salam International Center for Theoretical Physics, 34100 Trieste, Italy.}
\affiliation{$^4$ Department of Applied Physics, Faculty of Technology and Engineering,The Maharaja Sayajirao University of Baroda, Vadodara-390001, India.} 
\affiliation{$^5$ Department of Physics, University of Trieste, Strada Costiera 11, I-34151 Trieste, Italy.}
\affiliation{$^6$ CNR-IOM DEMOCRITOS Simulation Centre and SISSA, Via Bonomea 265, I-34136 Trieste, Italy.
}

\date{\today}

\begin{abstract}
We 
investigate the transport properties in out-of-equilibrium dynamics of strongly correlated dipolar fermions initially localized in one-dimensional inhomogeneous optical lattice. 
The 
dynamics is studied by 
experimentally measurable 
dynamical variables 
such as one-body density, pair-correlation function and size of the expanding cloud. In the noninteracting limit, we trace the usual fingerprints of ballistic expansion in the short time dynamics. However, dynamics is 
strongly affected by system size 
due to Pauli principle.
The dynamics also exhibits significant dependence on the sign of the interactions. 
We observe that strong repulsive dipolar interaction 
gives rise to many-body features in the dynamics, while strong attractive dipolar interaction leads to stabilized cluster states. 
Intermediate dipolar interaction are found to hinder expansion of correlations, 
while very strong dipolar interaction favors expansion of the cloud. Our work 
show the effect od 
dipolar interaction in the transport properties of interacting fermions, that can be 
studied in on-going experiments with ultracold dipolar fermions.  
\end{abstract}

\maketitle

\section{Introduction} \label{intro}
Transport properties is an active area of research in condensed matter and solid state physics. 
For strongly correlated systems, it is in general 
not straightforward to understand the transport properties even in models such as  the Hubbard model~\cite{Essler}. For real materials, additional difficulty arise due to the presence of impurities. 

Rapid advancement with ultracold atomic gases~\cite{RevModPhys.80.885,Lewenstein,Jaksch} and unprecedented control of system parameters in real time have allowed quantum simulation of the equilibrium properties of strongly correlated many-body systems~\cite{Esslinger,Hackermuller,Schneider:2008}. Ultracold atoms loaded in optical lattices offer the possibility to understand the expansion and transport phenomena in a clean and well-controlled environment. Using both bosons and fermions, it is possible to simulate strongly interacting quantum systems unveiling novel quantum phenomena~\cite{Bertini,Hild,Bulchandani,Schneider,Ronzheimer,Jie,Scherg,Chih-Chun,Strohmaier,Ott,Anderson,Peter,Nichols}. Although numerous efforts have been put forward to understand the interplay of strong correlations in determining the quantum many-body phases, non-equilibrium dynamics impose additional challenges and still remain under active study~\cite{Vasseur}.  

Expansion dynamics in low-dimensional quantum systems are of great interest from both theoretical and experimental perspectives due to intrinsic quantum correlation~\cite{Bertini}. It is well known that many-body physics in one dimension is significantly different from its higher dimensional counterparts. Both the dimension and interaction strength have 
a huge effect in the non-equilibrium transport phenomena~\cite{Schneider,Ronzheimer,Jie,Chih-Chun}. In the integrable limit, atoms expand ballistically, whereas interaction leads to diffusive dynamics, even smaller interactions suppress the expansion, typically leading to bimodal cloud shape. While the dynamics is independent of the sign of interaction, for strongly interacting bosons, a dimensional crossover from ballistic in one-dimension (1D) hard-core bosons to diffusive in two-dimension (2D) is observed~\cite{Schneider}. In the process of doubly or higher occupancy, the dynamics is more involved and include several intriguing phenomena like quantum distillation ~\cite{Scherg}. Several 
methods like density matrix renormalization group (DMRG)~\cite{Heidrich-Meisner} and multiconfigurational time-dependent Hartree for bosons (MCTDHB)~\cite{rhombik_scipost,rhombik_EPJPLUS} have been employed to study the expansion dynamics in one-dimension. 

The recent years have seen 
remarkable progress with atomic and molecular species exhibiting dipole-dipole interactions (DDI) such as dysprosium $^{161}$Dy~\cite{dys}, erbium $^{167}$Er~\cite{erb}, chromium $^{53}$C~\cite{cro}, potassium-rubidium
$^{40}$K$^{87}$Rb~\cite{Dipolar_mol}, sodium-lithium $^{23}$Na$^{6}$Li~\cite{Dipolar_mol1}, and sodium-potassium $^{23}$Na$^{40}$K
~\cite{Dipolar_mol2}. 
The dynamical properties of 
non-local interacting systems significantly differ from those
of short-range interacting ones in many aspects~\cite{Jens:2013,Schachenmayer:2013,Santos:2016,Anton:2016}. 
Such non-local interactions can exhibit novel features like; supersolids and dipolar supersolids~\cite{Li:2017, Boettcher:2019, Tanzi:2019, Tanzi:2019-2, Chomaz:2019, Natale:2019, Tanzi:2021}, dynamical phenomena such as time crystals~\cite{Kessler:2021, Kongkhambut:2022} and various Floquet phases~\cite{Bracamontes:2022, Sun:2023, Zhang:2023}. The out-of-equilibrium dynamics of isolated quantum systems with short-range interactions 
has been the subject of intense scrutiny, the presence of non-local and long-range interactions introduce more intriguing complexities~\cite{Defenu2021}.

The earlier studies of quantum transport of interacting bosons and fermions are mostly limited to the non-dipolar atoms, and the impact of the long-ranged dipolar interactions in the out-of-equilibrium expansion dynamics is yet to study. We follow the typical experimental protocol-dipolar fermions of variable interaction strengths, with the gas -- both with repulsive and attractive interactions -- is initially loaded in a combination of 1D optical lattice and strong harmonic confinement with different filling factor. In the out-of-equilibrium dynamics, the harmonic oscillator is eliminated suddenly and the dipolar fermions expand in the homogeneous optical lattice. We solve the time-dependent many-body Schr\"odinger equation for the continuum system and simulate the full many-body dynamics employing the multiconfigurational time-dependent Hartree method for indistinguishable components~\cite{Streltsov:2006,Streltsov:2007,Alon:2007,Alon:2008,Lode:2016,Fasshauer:2016}.

We consider a finite number of interacting dipolar fermions in a finite lattice of different filling factors covering entire range of weak to strong interactions, both repulsive and attractive. 
We study 
the one-body density matrix dynamics, the spreading of pair-correlation function and the size of the expanding cloud. Our observations are as follows. For noninteracting fermions, the initial two-body correlation in the prequench state is strictly determined by the number of fermions, as Pauli's principle is satisfied in the inhomogeneous lattice. In the post-quench dynamics, fermions first re-localize in the homogeneous lattice exhibiting intriguing features in the two-body density, while in the long-time, dynamics is driven by the density gradient only. For interacting fermions, we find a strong interplay of 
dipolar interaction. We observe identical density dynamics for both repulsive and attractive interactions when the interaction strength is very weak. However, a pronounced effect is observed for stronger interactions. When the strong repulsion manifests many-body features in different lattice sites, strong attraction leads to cluster states which remain highly stable in the entire dynamics. We also calculate the radius of the expanding cloud for long-time. The cloud size is maximum for noninteracting case, exhibiting a rapid spreading of cloud. However for intermediate interactions, the long-range repulsion in dipole-dipole interactions, hinders the spreading, resulting to smaller cloud size with periodic expansion and contraction. For very strong dipolar interactions, the cloud expands showcasing more chaotic density spreading towards outer lattice sites. To understand the interference effect of repulsive dipole-dipole interaction we calculate the reduced two-body density matrix for various interaction strengths. It exhibits how the correlation spreading becomes progressively stiffer for stronger dipolar interaction. This leads to formation of localized correlation around the middle lattices. Whereas much stronger interaction inhibits intriguing many-body features, correlation spreads reaching at the boundary.

The paper is organized as follows. In Sec. II, we introduce the model and the quench protocol. Sec. III presents the methodology and the dynamical quantities to be evaluated. In Sec. IV, we present the results for noninteracting fermions for different number of atoms loaded in the optical lattice. Sec. V presents the expansion dynamics for interacting cases ranging from strong repulsive to strong attractive dipolar interaction. Sec. VI concludes the summary.

\section{Hamiltonian and quench protocol}\label{numerics}
The equation of motion for $N$ interacting fermions is governed by the time-dependent Schr\"odinger equation as
$\hat{H} \psi = i \frac{\partial \psi}{\partial t}$ (with $\hbar=1$), with the many-body Hamiltonian
\begin{equation} 
\hat{H}(x_1,x_2, \dots x_N)= \sum_{i=1}^{N} \hat{h}(x_i) + \sum_{i<j=1}^{N}\hat{W}(x_i - x_j).
\label{propagation_eq}
\end{equation}

\begin{figure}
\centering
\includegraphics[width=0.9\linewidth]{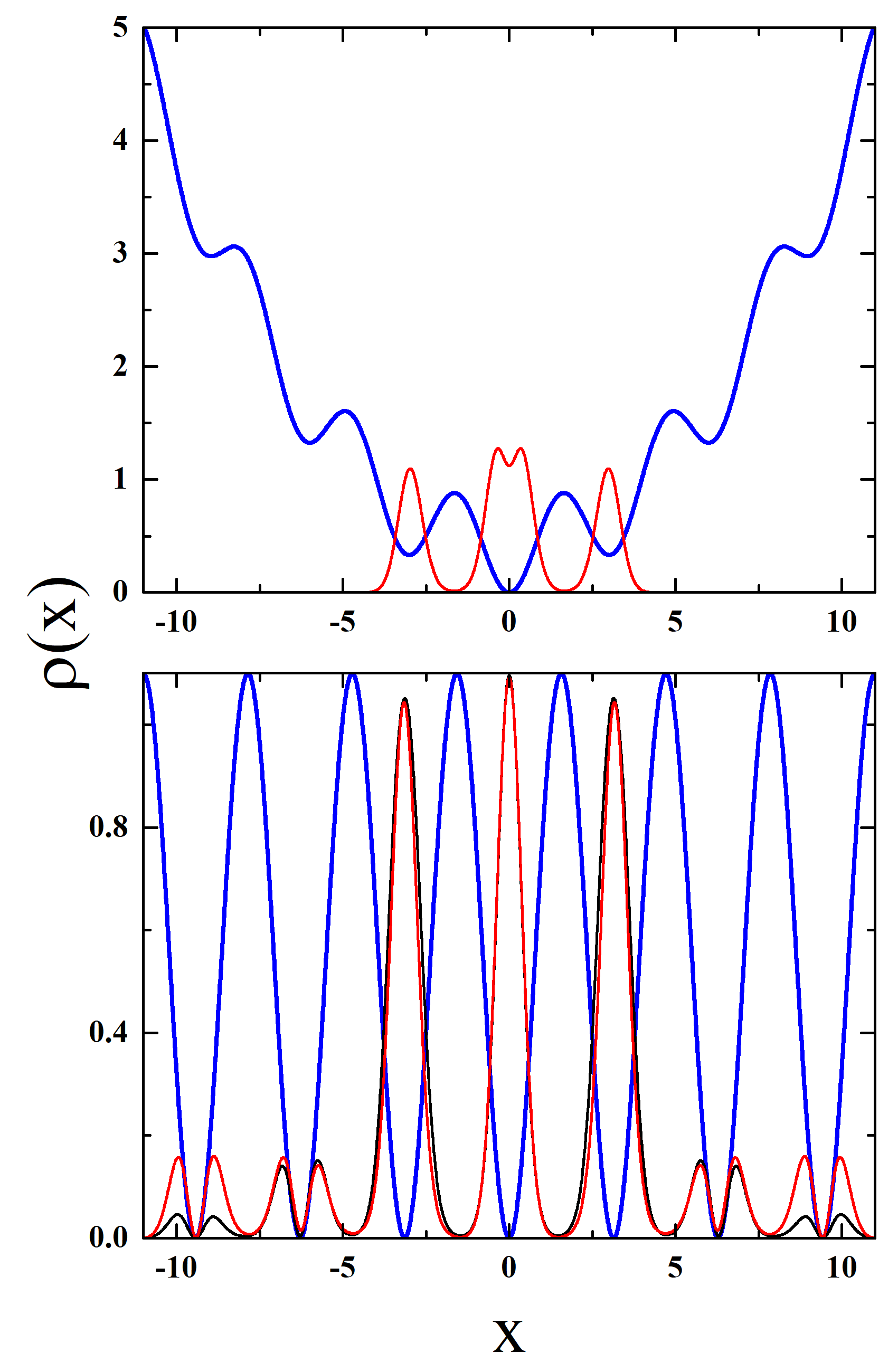}
\caption{Expansion of fermionic atoms after a quench of the trapping potential. Top panel: Initial density profile (red curve) with $N=4$ noninteracting fermions in the combination of optical lattice of lattice depth $V_0=9 E_r$, lattice size with $S=7$ sites and a strong harmonic trap with $\omega^2=0.4 E_r^2 {\hbar}^{-2}$. The blue curve represents the effective trap in a reduced scale. Bottom panel: Density of expanding fermionic atoms after sudden removal of the harmonic confinement and cloud expands in the homogeneous lattice. Black curve ($t=10.0$) and red curve ($t=50$),  present expansion of fermions to outer lattice. The blue curve is the homogeneous lattice in reduced scale for the postquench state. Computation is done with $M=5$ orbitals, see the text for discussion.
}
\label{fig1}
\end{figure}

The one-body part of the Hamiltonian is $\hat{h}(x) = \hat{T}(x) + \hat{V}_{OL}(x)$, where $\hat{T}(x) = -\frac{\hbar^2}{2m} \frac{\partial^2}{\partial x^2}$ is the kinetic energy operator and $\hat{V}_{OL}(x) = V_0 \sin^2(kx)$.
$V_0$ is the lattice depth and $k$ is the wave vector. 

The two-body interactions $\hat{W}(x_i - x_j)$ appearing in Eq.~(1) are modeled with long-range dipole-dipole interactions $\hat{W_D}(x_i-x_j) = \frac{g_d}{|x_i-x_j|^3 + \alpha}$.
Here, $g_d$ controls the strength of the DDI and it is positive for repulsive interactions and negative for attractive interactions.  The parameter $\alpha$ denotes a renormalization factor that arises when DDI are confined to a one-dimensional geometry. We keep it constant at $\alpha=0.05$ for all our computations.

With the above Hamiltonian we prepare different setups with $N=4,5,6,7$ number of interacting fermions which are initially trapped in a combination of 1D optical lattice with a strong harmonic confinement, $V_\text{trap}=\frac{1}{2} \omega^2 x^2 + V _0 \sin^2 (k x)$. The spatial extension of the optical lattice is $x_{\max}=7 \pi/2$ to $x_{\min} = - 7 \pi/2$, which restricts the number of lattice sites to 7 and is kept fixed in the entire simulation. The lattice depth $V_0= 9 E_r$ is also kept fixed throughout our calculation, where $E_r = \frac{\hbar^2 k^2}{2 m}$ is the lattice recoil energy. The many-body Hamiltonian is rescaled in the unit of $E_r$. Thus, the units of time, length, and interaction strength are $\hbar E_r^{-1}$, $k^{-1}$, and $2 E_r k^{-1}$, respectively. We consider both repulsive and attractive dipolar fermions in the range $g_d \in [-1.0 E_r, 1.0 E_r]$.  We have probed a fine array in this range, but print results for only some selective cases to be concise. The initial state is prepared with noninteracting or interacting fermions in the effective trap with frequency $\omega^2=0.4 E_r^2 {\hbar}^{-2}$. In the out-of-equilibrium dynamics, the harmonic oscillator trap is suddenly eliminated, thus transporting the fermions in the homogeneous lattice.

In Fig.~\ref{fig1}(top), we plot the initial state density for $N=4$, noninteracting fermions in the combination of lattice and harmonic trap. The initial state is effectively confined in three sites due to strong transverse confinement. The central lattice traps two fermions, the dip in the density signatures the Pauli principle to be satisfied. The other two fermions reside in the left and right wells. In Fig.~\ref{fig1}(bottom), we present the density of the postquench state for some specific time points. In the postquench dynamics, the harmonic trap is suddenly switched off and fermions expand. Note that in absence of any external gradient, the fermionic transport is absolutely determined by the density gradient. As the fermions are transported to the outer lattices, the dip in the central density disappears. Even in the long-time dynamics, the fermions are mostly confined in the central and middle wells, and only a small fraction transports in the outer wells. As the fermions are noninteracting, the postquench transport manifests only the dilution of fermion mass. We observe significant effect of long-range dipolar interaction with intriguing many-body features in the density dynamics of interacting fermionic transport as discussed later.

\section{Methods}
\label{sec:methods}
To investigate the dynamics of dipolar fermions under the sudden quench, we employ the multiconfigurational time-dependent Hartree method for indistinguishable particles~\cite{Streltsov:2006, Streltsov:2007, Alon:2007, Alon:2008} implemented by the MCTDH-X software~\cite{Alon:2008,Lode:2016,Fasshauer:2016,Lin:2020,Lode:2020,MCTDHX}. MCTDH-X relies on a time-dependent variational optimization procedure in which the many-body wavefunction is decomposed into an adaptive basis set of $M$ time-dependent single-particle functions called orbitals. MCTDH-X has been widely used to study ground-state and dynamical properties of long-range interacting systems~\cite{fischer,Bera:2019,rhombik_EPJPLUS,chatterjee:2018,chatterjee:2019,chatterjee:2020,Hughes:2023}.

MCTDH-X solves the many-body Schr\"{o}dinger equation by recasting the many-body wave function as an adaptive superposition of $M$ time-dependent permanents constructed from single-particle wave functions, called orbitals. 
Both the coefficients and the basis functions in this superposition are optimized in time to yield either ground-state information (via imaginary time propagation) or full-time dynamics (via real-time propagation).
The ansatz for the many-body wave function is the linear combination of time dependent permanents
\begin{equation}
\vert \Psi(t)\rangle = \sum_{\bar{n}}^{} C_{\bar{n}}(t)\vert \bar{n};t\rangle,
\label{many_body_wf}
\end{equation}
The vector $\vec{n} = (n_1,n_2, \dots ,n_M)$ represents the occupation of the orbitals, with the constraint that $n_1 + n_2 + \dots +n_M = N$, which ensures the preservation of the total number of particles. 
Distributing $N$ spin-polarized fermions over $M$ time dependent orbitals, the number of permanents become roughly $ \left(\begin{array}{c} M \\ N \end{array}\right)$.
The permanents are constructed over $M$ time-dependent single-particle wave functions, called orbitals, as 
\begin{equation}
\vert \bar{n};t\rangle = \prod^M_{k=1}\left[ \frac{(\hat{b}_k^\dagger(t))^{n_k}}{\sqrt{n_k!}}\right] |0\rangle 
\label{many_body_wf_2}
\end{equation}

Where $|0\rangle$ is the vacuum state and $\hat{b}_k^\dagger(t)$ denotes the time-dependent operator that creates one boson in the $k$-th working orbital.

It is important to note that both the expansion coefficients $\left \{C_{\bar{n}}(t)\right\}$ and the working orbitals $\psi_i(x,t)$ that are used to construct the permanents $\vert \bar{n};t\rangle$ are fully variationally optimized and time-dependent quantities~\cite{TDVM81,variational1,variational3,variational4}. The time-dependent optimization of the orbitals and expansion coefficients ensure that the MCTDH-X can accurately describe the dynamics of strongly interacting atoms. MCTDH-X has found extensive application in the study of dynamics in various trap geometries and interaction strengths~\cite{rhombik_pra,rhombik_jpb,rhombik_aipconference,rhombik_pre,rhombik_scipost}. As the permanents are time-dependent, a given degree of accuracy is achieved with shorter time compared to a time-independent basis.
The time-adaptive many-body basis set employed in MCTDH-X allows for the dynamic tracking of correlations that arise from inter-particle interactions. 

Utilizing several orbitals in the many-body ansatz allows us to capture the fully correlated many-body states. MCTDH-X ansatz becomes exact in the limit $M \rightarrow \infty$. However, for numerical simulation, often a finite number of orbitals is enough to accurately describe short-to-medium dynamics. The exactness of numerical simulation is established by orbital convergence. MCTDH-X simulation with $M$ orbitals becomes exact when the population of an additive orbital becomes insignificant and dynamical observables becomes converged. To simulate the $N$-body interacting fermion systems we require $M \ge N + 1$  orbitals, in principle $M=N$ would be sufficient, however it is often degenerate. For strongly interacting systems, a larger number of orbitals might be required
to capture many-body correlation which also exhibits to negligible population in the last orbitals for the entire time dynamics. 

To extract information from the system, we calculate several observables from the many-body state $\left| \Psi(t) \right>$.
To probe the spatial distribution, we calculate the one-body density as
\begin{align}
\rho(x; t) &= \left< \Psi(t) \right| \hat{\Psi}^{\dagger}(x) \hat{\Psi}(x) \left| \Psi(t) \right>.
\end{align}

To measure the degree of correlation, we calculate the reduced two-body density matrix, defined as \cite{sakmann}
\begin{equation}
\rho^{(2)}(x,x'; t) = \left< \Psi(t) \right| \hat{\Psi}^{\dagger}(x) \hat{\Psi}^{\dagger}(x') \hat{\Psi}(x') \hat{\Psi}(x) \left| \Psi(t) \right>.
\end{equation}
The diagonal of the two-body reduced density matrix is also equivalent to the pair correlation function. 
The population in each orbital is defined as the eigenvalue of the reduced one-body density matrix
\begin{align}
\rho^{(1)}(x,x') &=  \left< \Psi\right| \hat{\Psi}^{\dagger}(x') \hat{\Psi}(x) \left| \Psi \right> \\ \nonumber
                  &= \sum_i \rho_i \left(\phi_{i}^{(\mathrm{NO})}(x')\right)^{*} \phi_{i}^{(\mathrm{NO})}(x).\label{eq:RDM1}
\end{align}
$\phi_i^{(\mathrm{NO})}$  are termed as natural orbitals and $\rho_i$ is the population in each orbital.

\section{Expansion dynamics for noninteracting fermions}

The seminal work on fermionic transport of non-dipolar atoms in a homogeneous Hubbard model in 2D has exhibited ballistic transport of noninteracting fermions when the initially confined atomic cloud expands suddenly~\cite{Schneider}. Whereas small interactions can drastically reduce the mass transport, can change the shape of expanding cloud. However the situation may be more involved in one dimension and has not been explored yet how the transport mechanism will be affected by the long-range dipolar interaction. 

\begin{figure}
\centering
\includegraphics[scale=0.1, angle=270]{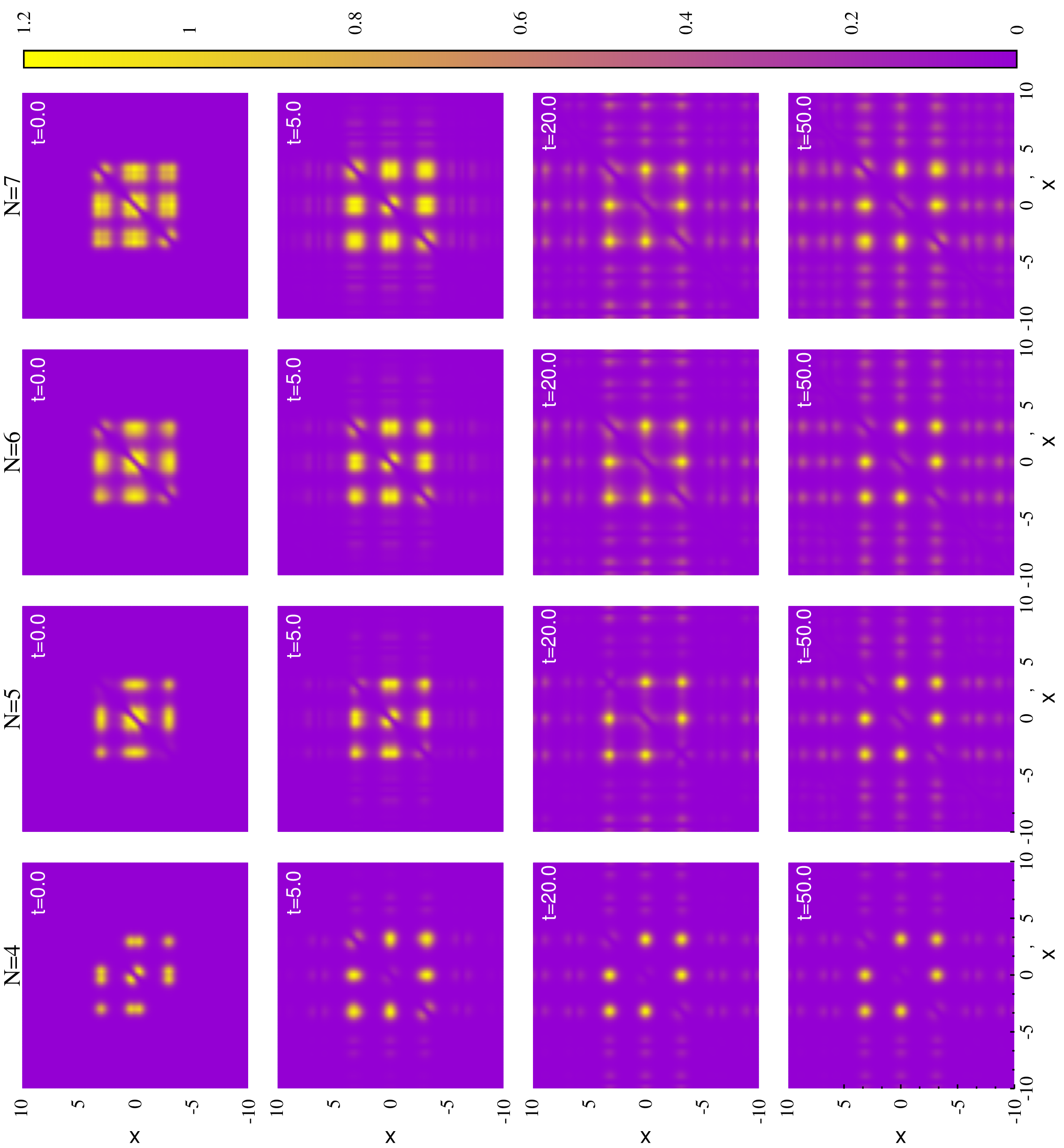}
\caption{Dynamics of the reduced two-body density matrix $\rho^{(2)}(x,x^{\prime},t)$ post quench for noninteracting fermions with different system size; $N=4,5,6,7$. The lattice size is kept fixed to $S=7$ sites, lattice depth is $V_0=9 E_r$. Computation is done with $M=N+1$ orbitals.  See the text for discussion.
}
\label{fig2}
\end{figure}

In this section we specifically focus on the transport of noninteracting fermions in the postquench dynamics for various $N$, keeping lattice size, lattice depth fixed. Initially we follow the set up presented in Fig.~\ref{fig1}, dipolar fermions of various filling factor are trapped in the inhomogeneous lattice created by the combination of 1D optical lattice and strong harmonic confinement. 
In the postquench dynamics, the oscillator is removed and the fermions expand in the homogeneous lattice. It should be believed that noninteracting fermions in 1D will exhibit ballistic transport. However, we observe some additional features in the density dynamics as the fermions must re-localize to satisfy the Pauli principle, which is best represented by the reduced two-body density matrix (2-RDM) as given in Eq.~(5). This is equivalent to the pair correlation function that describes the probability of finding one fermion at $x$ when the other fermion is located at $x^{\prime}$. In Fig.~\ref{fig2}, we present $\rho^{(2)}(x,x^{\prime};t)$ for selected time points, $t=0.0, 5.0, 20.0, 50.0$. Initially the two-body correlation strongly depends on the filling factor, however in the long-time dynamics they become identical as the postquench dynamics is dictated by the dilution of the systems only. 

For $N=4$, at $t=0.0$, two bright lobes with clear diagonal extinction (correlation hole) in the central lattice exhibit the presence of two fermions satisfying the Pauli principle. The extinct diagonal indicates that two fermions can not be found in the same place, $x=x^{\prime}$.  For $N=5$, the lobes become brighter only as the central lattice now confine three fermions; however, the correlation hole along the diagonal is maintained. For $N>5$, when the same physics is explained for the central lattice, two other pairs of less bright lobes with diagonal extinction appear in the left and right wells, signifying the presence of more than one fermion in the outer sites. These initial correlations can also be verified from the density signature as presented in the next section (Fig.~\ref{fig4}). 

At $t=5.0$, when the fermions transport in the homogeneous lattice, we observe distinct features. For $N=4$, when the central pair of bright lobes disappears, new pairs of lobes with correlation hole but of diminished intensity become visible in the left and right well. Thus the pair of fermions, initially trapped in the central lattice transports to outer lattice. For $N=5$, the central lobes do not disappear, only become diminished, the trapped fermions partially transport to outer sites, generating new lobes. For $N=6$ and $N=7$, we observe three pairs of lobes, each with clear correlation hole and of almost same intensity in the three middle lattice sites. Thus, just after quench, the fermions mostly rearrange  within the three wells for all sets of particle numbers. For longer time, $t=20$ and $50$, the pair-correlation becomes independent of particle number and spreads at the boundary. We conclude that the noninteracting fermions demonstrate stringent features in the short-time transport, however in the asymptotic limit, the system becomes dilute only.

\begin{figure}
\centering
\includegraphics[width=\linewidth]{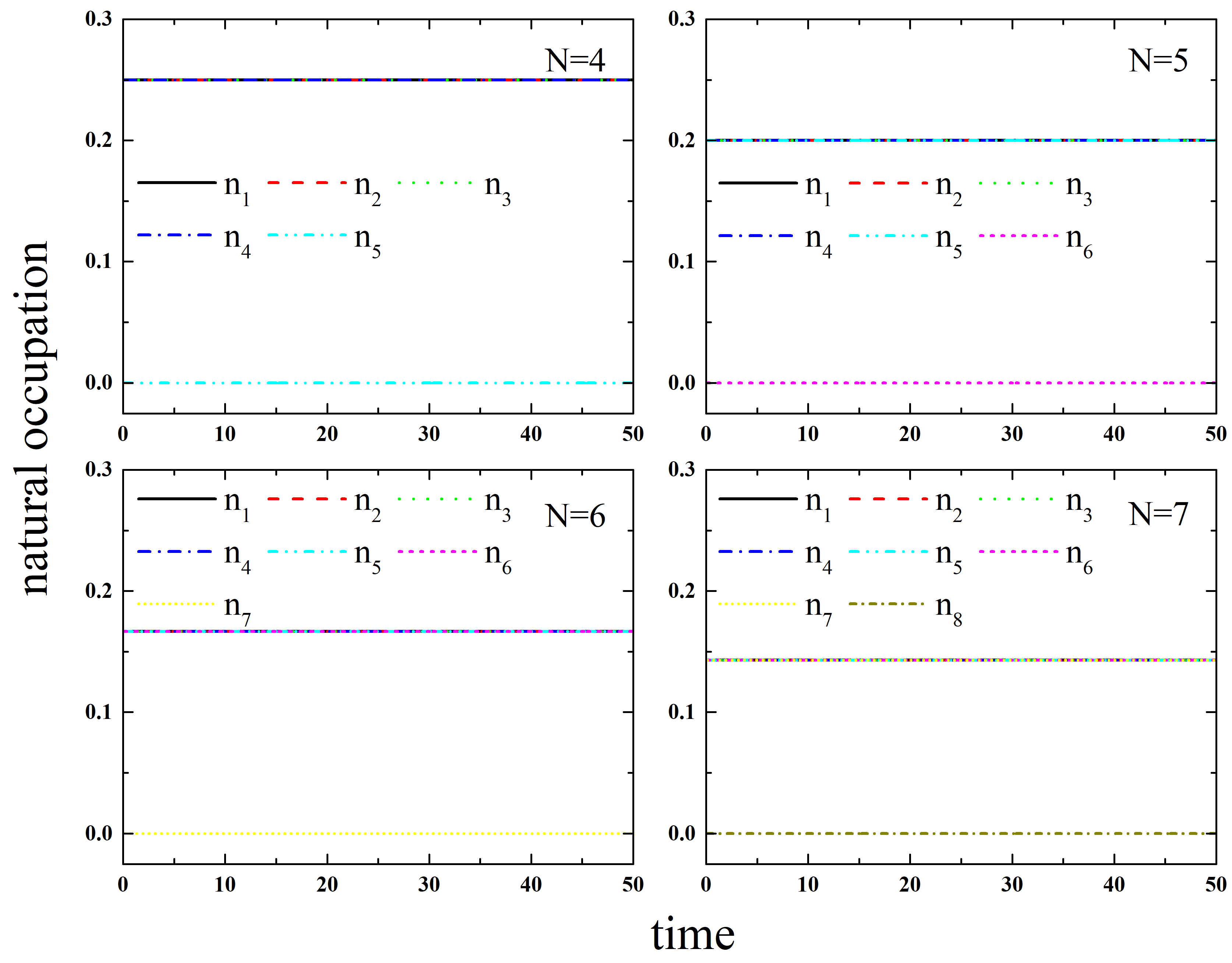}
\caption{ Time evolution of orbital occupation in the post quench dynamics for different number $(N)$ of noninteracting fermions. Computation is done with $M=N+1$ orbitals, each contributing $\frac{1}{N}$ of the total population. As the lowest $M=N$ orbitals contribute equally, the plots overlap, forming a continuous line. See the text for discussion.
}
\label{fig3}
\end{figure}

Orbital convergence is also an important issue to clearly exhibit numerically exact solutions. Fig.~\ref{fig3}, shows the time evolution of the occupation in natural orbitals for post quench dynamics with noninteracting fermions. For each system size $(N)$, computation is done with $M=N+1$ orbitals. In all cases, it is uniquely established that the number of dominating orbitals is clearly the number of fermions exhibiting the Pauli principle, each fermion resides in a single orbital. For $N=4$, the first four natural orbitals contribute equally $\simeq$ 25 \%; for $N=5$, each of five orbitals contribute equally $ 20 \%$; for $N=6$, six orbitals contribute, each orbital exhibit 16.66 \% occupation; for $N=7$, the contribution of each $M=7$ orbitals is $14.28 \%$ .

\section{Expansion dynamics for interacting fermions}
\subsection{Initial density signatures of interacting fermions}
\begin{figure}
\centering
\includegraphics[width=\linewidth]{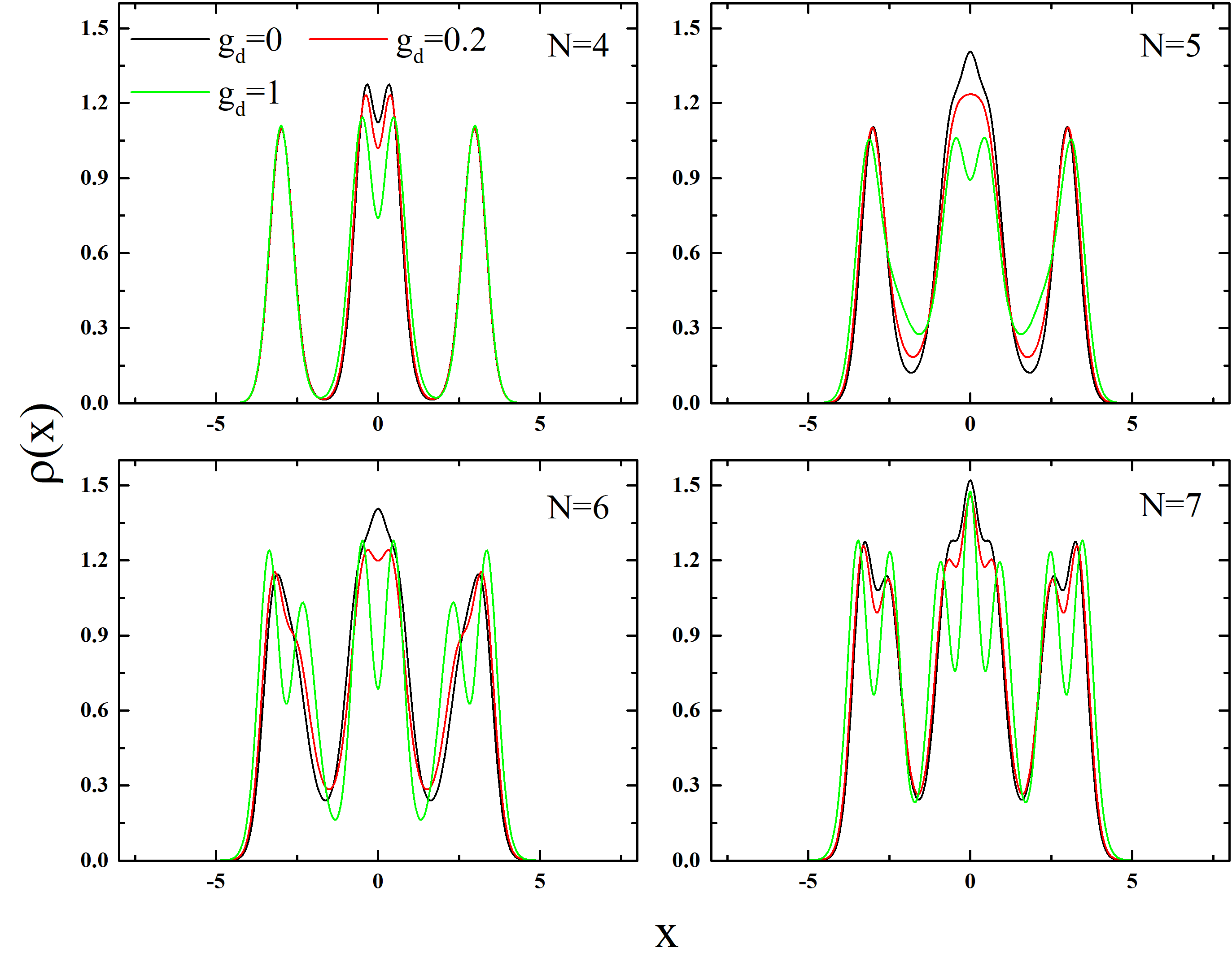}
\caption{ Initial density profiles for interacting fermions in the combination of optical lattice of $S=7$ sites, lattice depth $V_0=9 E_r$, and harmonic confinement with frequency $\omega=0.4 E_r^2 {\hbar}^{-2}$. 
For all cases, the density is confined in the three middle wells. With increase in interaction strength $g_d$, the density becomes modulated with clear signature of humps. For stronger interaction, the number of distinct humps in the density gradually approaches to the number of fermions satisfying Pauli principle. For noninteracting fermions, computation is done with $M=N+1$ orbitals. However for interacting cases, $M=12$ orbitals are required for systems with $N=4$ and $5$ fermions. Systems with larger filling factor, $N=6$ and $7$ fermions, we utilize $M=10$ orbitals. See the text for discussion.
}
\label{fig4}
\end{figure}

Fig.~\ref{fig4} represents the ground state density in the initial setups in the combination of optical lattice and harmonic oscillator for interacting fermions with $N=4,5,6,7$. We present some representative cases of interaction strength, $g_d=0.0, 0.2, 1.0$. For all choices of $N$, increasing $g_d$ gradually from noninteracting to weak and strong interactions, fermions redistribute in the lattice sites. The kink in the density signifies the Pauli exclusion principle to be satisfied by two neighboring bosons sitting in the same lattice. With increase in interaction, due to extra repulsion arising from long-range repulsive tail, the kinks become deep and prominent. 

For $N=4$, in the noninteracting limit, the dip in the central lattice exhibits two fermions reside in the central well, the other two fermions are trapped in the left and right wells. With increase in $g_d$, the dip in the central lattice becomes more prominent signifying that two fermions feel strong repulsion, leaving the other two fermions in the outer lattices unaffected.  For $N=5$, with $g_d=0$, three fermions reside in the central lattice, exhibiting by one distinct peak with two pseudo kinks. Two other fermions reside in the left and right wells. With increase in interaction, the fermions reshuffle their position, however, even for $g_d=1.0$, five distinct peaks are not observed. The calculation is repeated with higher orbitals to rule out the convergence issue. For $N=6$, in the non-interacting limit, six fermions are distributed in the three wells, with clear dip or pseudo-kink structures explaining the Pauli principle satisfied in each lattice site. With increase in $g_d$, the distribution of fermions becomes uniform only. For $N=7$ with $g_d=0$, three bosons settle in the central lattice, left and right wells trap a pair of fermions. With increase in interaction, seven prominent peaks signify the distribution of the strongly interacting fermions.  It is to be noted that for all cases of filling factor and interaction strength, the initial density is confined within the three middle wells due to strong harmonic confinement. 

We find that orbital convergence becomes challenging for stronger interaction and with lower filling factors; we need to utilize $M=12$ orbitals for computation when the number of interacting fermions is $N=4$ and $5$. With $N=6$ and $N=7$, when the filling factor is close to one, for the entire choices of interaction strength, $M=N+1$ orbitals should be sufficient, however, we utilize $M=10$ orbitals to establish convergence. For noninteracting cases, we utilize $M=N+1$ orbitals for all cases of filling factor. The details of orbital convergence is presented in Appendix A.

\begin{figure}
\centering
\includegraphics[scale=0.12, angle=-90]{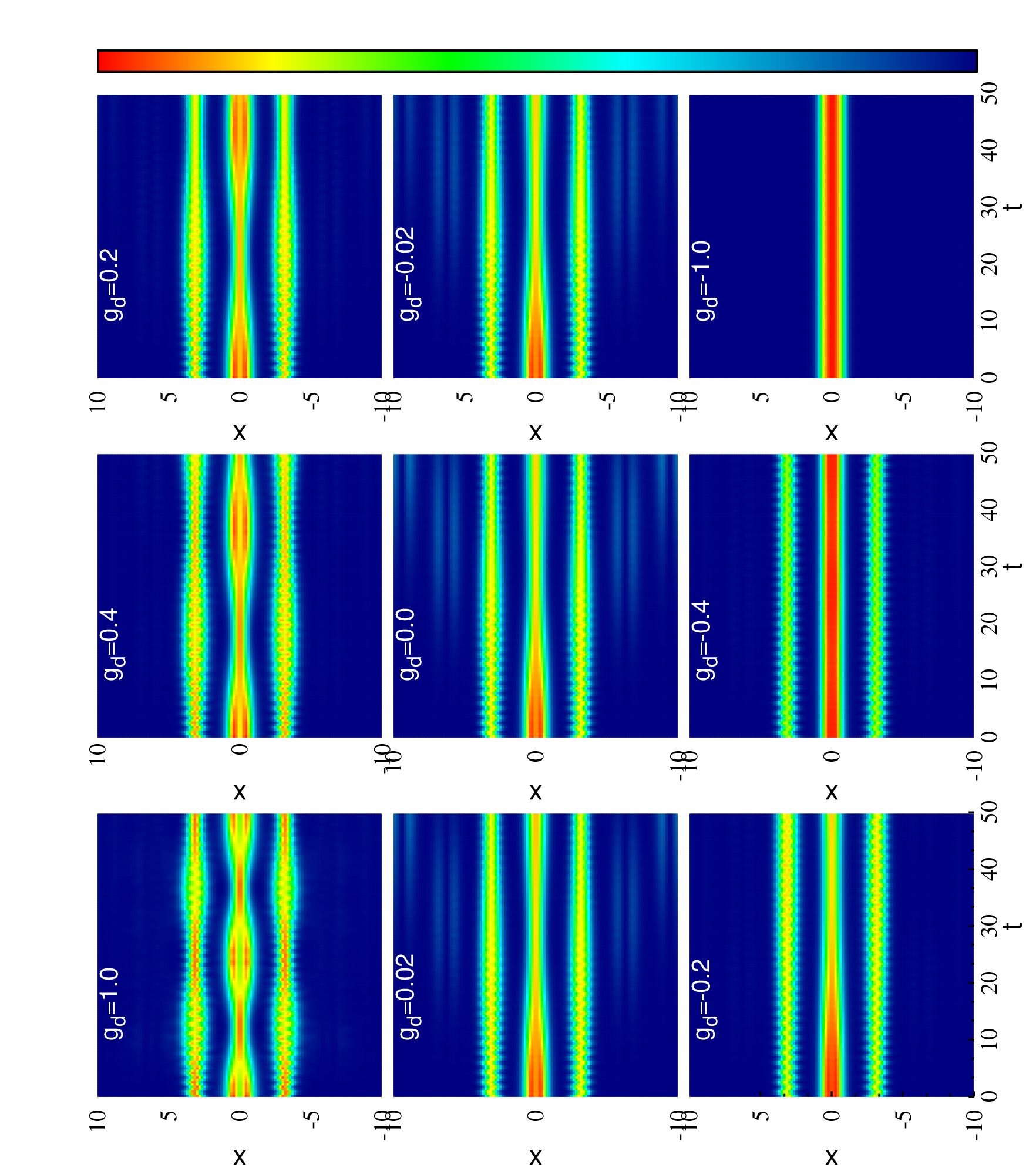}
\caption{ Dynamics of reduced one-body density $\rho(x,t)$ post quench for $N=4$ fermions for different interactions. For weak interaction, the evolution of the density is identical for positive and negative $g_d$ values. However for stronger $g_d$, when repulsive interaction manifests many-body features, attractive interaction leads to highly stable cluster formations. The computation is done with $M=5$ orbitals for $g_d=0.0,0.02$, whereas $M=12$ orbitals are needed for $g_d=0.2,0.4,1.0$. For attractive interaction $M=8$ orbitals are used. See the text for discussion.
}
\label{fig5}
\end{figure}

\subsection{One-body density dynamics}

In Fig.~\ref{fig5}, we plot the one-body density dynamics $\rho(x,t)$ of $N=4$ interacting fermions for nine representative values of $g_d$ that includes noninteracting, repulsive and attractive interactions showcasing the appearance of different dynamical behaviors. For all $g_d$ values, the prequench states remain confined almost in three lattice sites. For noninteracting and weakly interacting fermions $g_d=0.02$ (repulsive as well as attractive), the density dynamics are almost indistinguishable. At $t=0$, the two bright spots in the central lattice signify the presence of two fermions. In the post quench dynamics, two bright jets are visible till $t \simeq 15$. After that, the jets gradually disappear signifying that fermions from the central lattice transport to outer lattice sites. When the density in the central lattice shrinks, that in the outer lattice expands configuring the fermionic transport. At much longer time, a halo structure of low-density cloud appears around the dense central cloud. It signifies the transport of fermions to the lattice boundaries. For such weak interactions, long-range tail in the dipole-dipole interaction does not play any profound role and in the asymptotic limit the fermions simply transport to outer lattice sites.

For slightly stronger interaction, $g_d=0.2$, the central lattice exhibits the breathing dynamics, the cloud expands and contracts almost in  a periodic manner. The density dynamics in the outer lattice sites does not exhibit any many-body features; when the density shrinks in the central lattice, it expands in the outer lattice only. The low-density halo cloud is not visible anymore, signifying that long-range repulsive tail prohibits the fermions to reach lattice boundary. For attractive interaction with $g_d=-0.2$, the density is significantly localized in the central lattice without any distinction of interacting fermions, the density in the outer lattice does not exhibit any significant feature. The postquench dynamics remains stable.

With further increase in interaction strength, $g_d=0.4$, the breathing dynamics in the central lattice becomes more prominent, the density in the outer lattice sites also develop some features of expansion and contraction in the opposite time scale of central lattice dynamics. In the course of time, two fermions either settle in or travel away from the central lattice with a fixed time period, exhibiting fermionic oscillation. For $g_d=-0.4$, the fermions become simply localized, forming a cluster state in the central lattice, which remain highly stable in the entire dynamics.

For very strong interaction, $g_d$=1.0, the time period of breathing oscillation in the central lattice is significantly reduced, when the clear expansion-contraction dynamics is also developed in the outer lattice.  The stringent many-body features are essentially confined in the three middle wells and even very strong interaction does not allow the cloud expansion to the outer lattice. In contrast, for $g_d=-1.0$, the strongly attractive dipolar fermions form a self bound soliton with negative energy in the central lattice. It does not exhibit any kind of destabilization in the quench process. 

The one-body density dynamics, for other higher filling factors with $N=6$ and $N=7$ and varying interaction strengths are presented in the Appendix B. For larger filling factor, we observe that very strong interaction essentially facilitates expansion to lattice boundary unlike the low filling case.

\begin{figure}
\centering
\includegraphics[width=\linewidth]{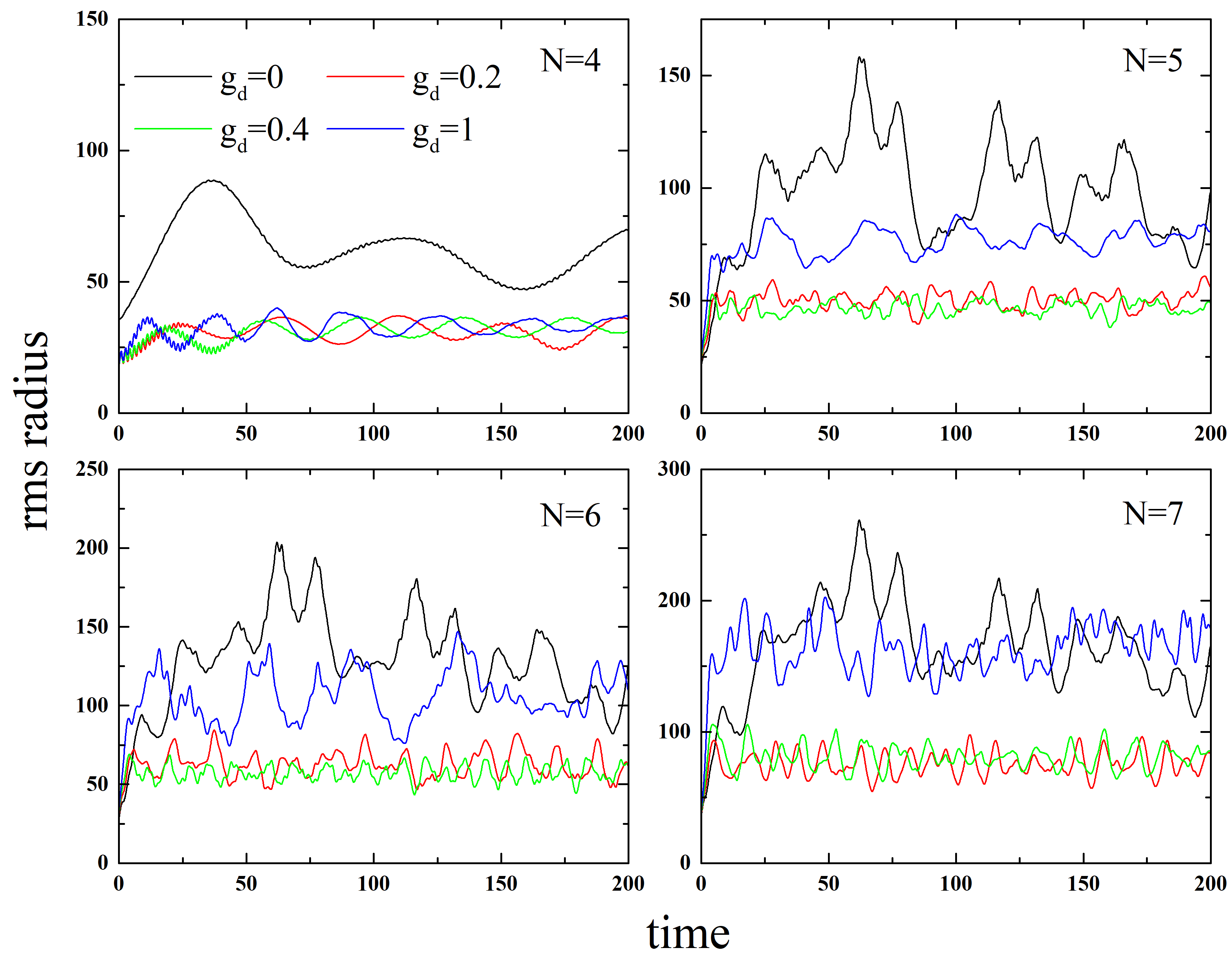}
\caption{ Time evolution of root mean square (rms) radius of the fermionic cloud for different system sizes and interaction strengths. For noninteracting cases, the fermions simply expand freely. For stronger dipolar interaction, $g_d$=0.2 and 0.4, the spreading is arrested for all system sizes and reveal breathing dynamics. Whereas very strong dipolar interaction enhances the spreading and cloud size increases chaotically. See the text for discussion.
}
\label{fig6}
\end{figure}

The corresponding root mean square (rms) radius calculated
from $\sqrt{\int x^2 \rho(x) dx}$ determines the average size of the expanding cloud and in presented in Fig.~\ref{fig6} for different system sizes. We portrait time evolution of rms radius for different filling factors with $g_d=0.0, 0.2,0.4,1.0$, for long time, $t=200$. 

For $g_d=0.0$, when the fermions transport due to their density gradient, the cloud size becomes bigger with increase in system size. Initially, we observe the ballistic expansion, however in the long-time some fluctuating peaks appear. For $N=4$, the cloud size exhibits some regular peaks, manifesting that the expanding cloud reaches to boundary and then returns. For higher filling factors, the cloud expands very fast, hits the boundary and reflects back which results to fluctuations in the cloud size.

For interacting fermions, we observe strong interplay of the repulsive long-range tail in the dipolar interaction. For $g_d=0.2$ and $0.4$, independent of system size (filling factor), the cloud expansion is strongly reduced. It exhibits that dipolar interaction hinders the cloud expansion for stronger interaction. The oscillatory behaviors exhibit the breathing dynamics of the expanding cloud. The average cloud size is $\simeq 25$ for $N=4$; for $N=5$ and $6$ cloud size is $\simeq 50$; whereas the cloud size is significantly bigger $\simeq 90$ for $N=7$. For very strong interaction, $g_d=1.0$, long-range interaction suddenly enhances the expansion. For $N=4$, the effect of strong long-range interaction is still not significant, the cloud size remains as before, only the time period of breathing dynamics is significantly reduced as observed in Fig.~\ref{fig5}. However, for the rest of system sizes, the effect of strong long-range repulsion is gradually enhanced with increase in system size. For $N=7$, we observe that cloud size almost becomes double due to profound effect of long-range repulsion. We also measure the cloud size in the larger lattice and conclude the same observation (see the Appendix C).

\subsection{Pair-correlation dynamics for unit filling}

\begin{figure}
\centering
\includegraphics[scale=0.2, angle=-90]{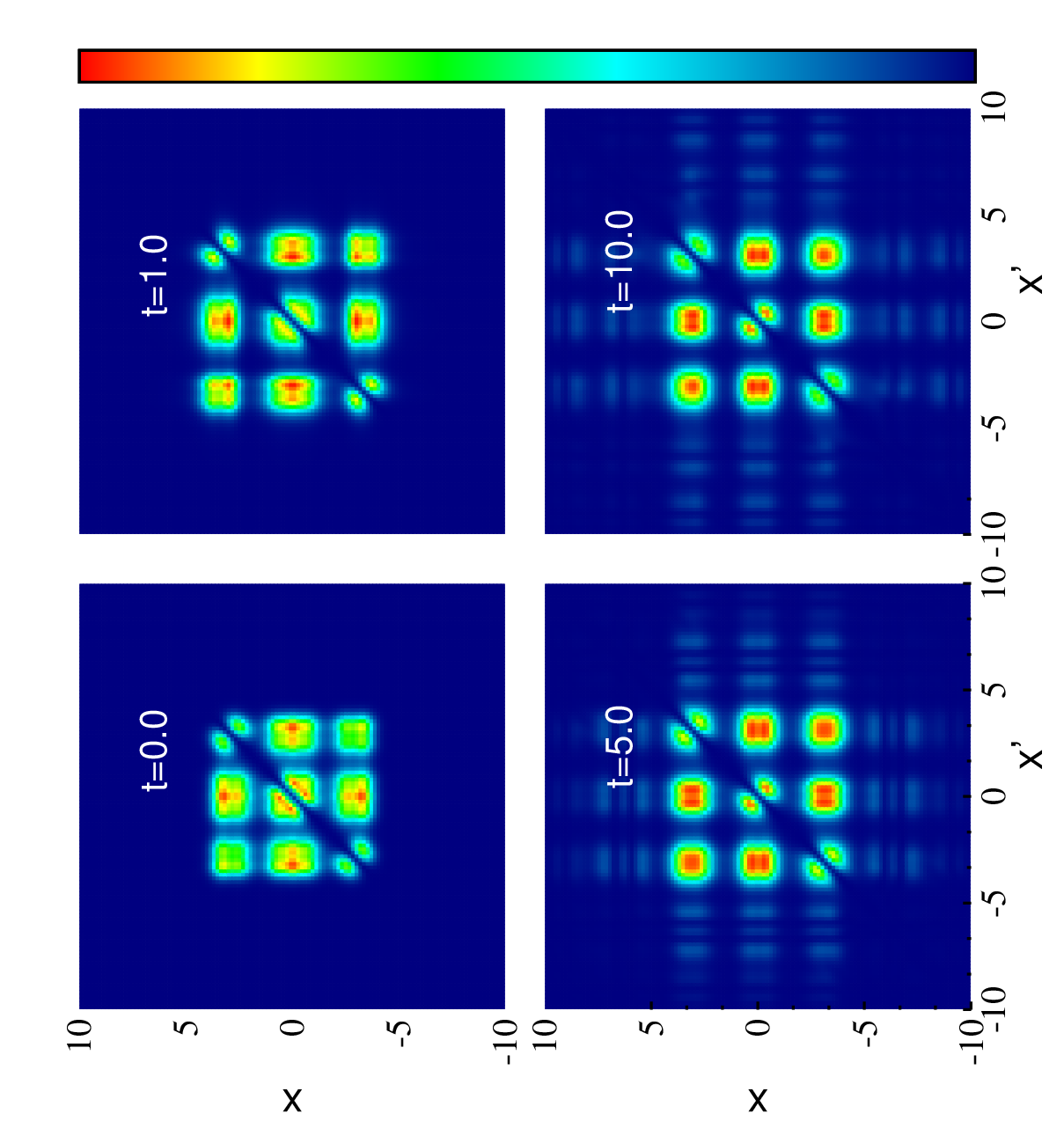}
\caption{ Dynamics of the reduced two-body density matrix $\rho^{(2)}(x,x^{\prime})$ post quench for interacting fermions with unit filling factor, $N=7$ fermions expands in $S=7$ lattice sites. The interaction strength is $g_d=0.02$, the weak dipolar interaction enhances expansion, however the core of fermionic cloud is significantly confined in three central lattice sites. Computation is done with $M=10$ orbitals.  See the text for discussion.
}
\label{fig7}
\end{figure}

\begin{figure}
\centering
\includegraphics[scale=0.2, angle=-90]{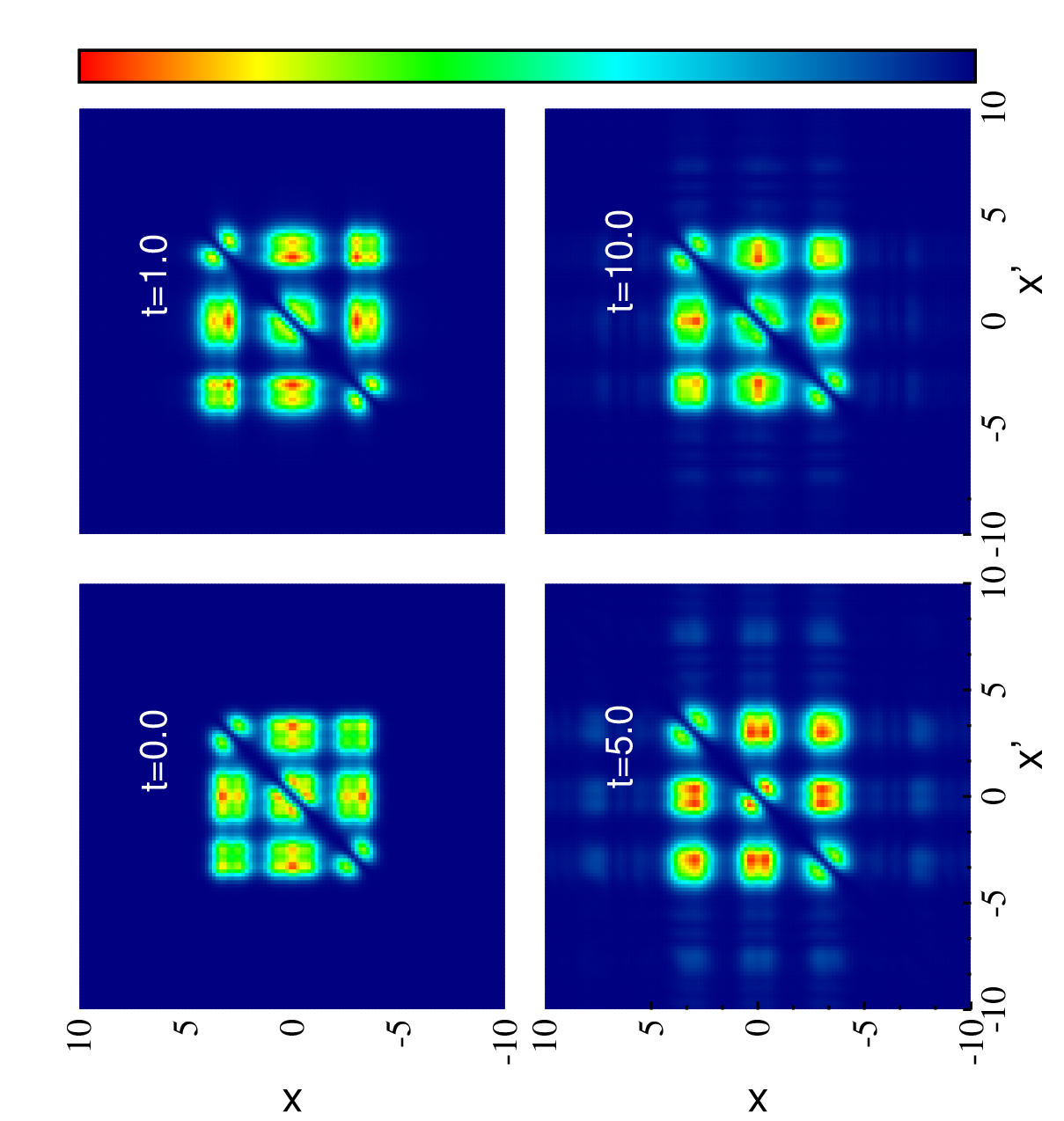}
\caption{ Dynamics of the reduced two-body density matrix $\rho^{(2)}(x,x^{\prime})$ post quench for interacting fermions with unit filling factor, $N=7$ fermions expands in $S=7$ lattice sites. The interaction strength is $g_d=0.2$, the stronger dipolar interaction initially enhances expansion, however it is gradually arrested with time.  The many-body features of the core cloud does not change significantly. Computation is done with $M=10$ orbitals.  See the text for discussion.
}
\label{fig8}
\end{figure}

\begin{figure}
\centering
\includegraphics[scale=0.2, angle=-90]{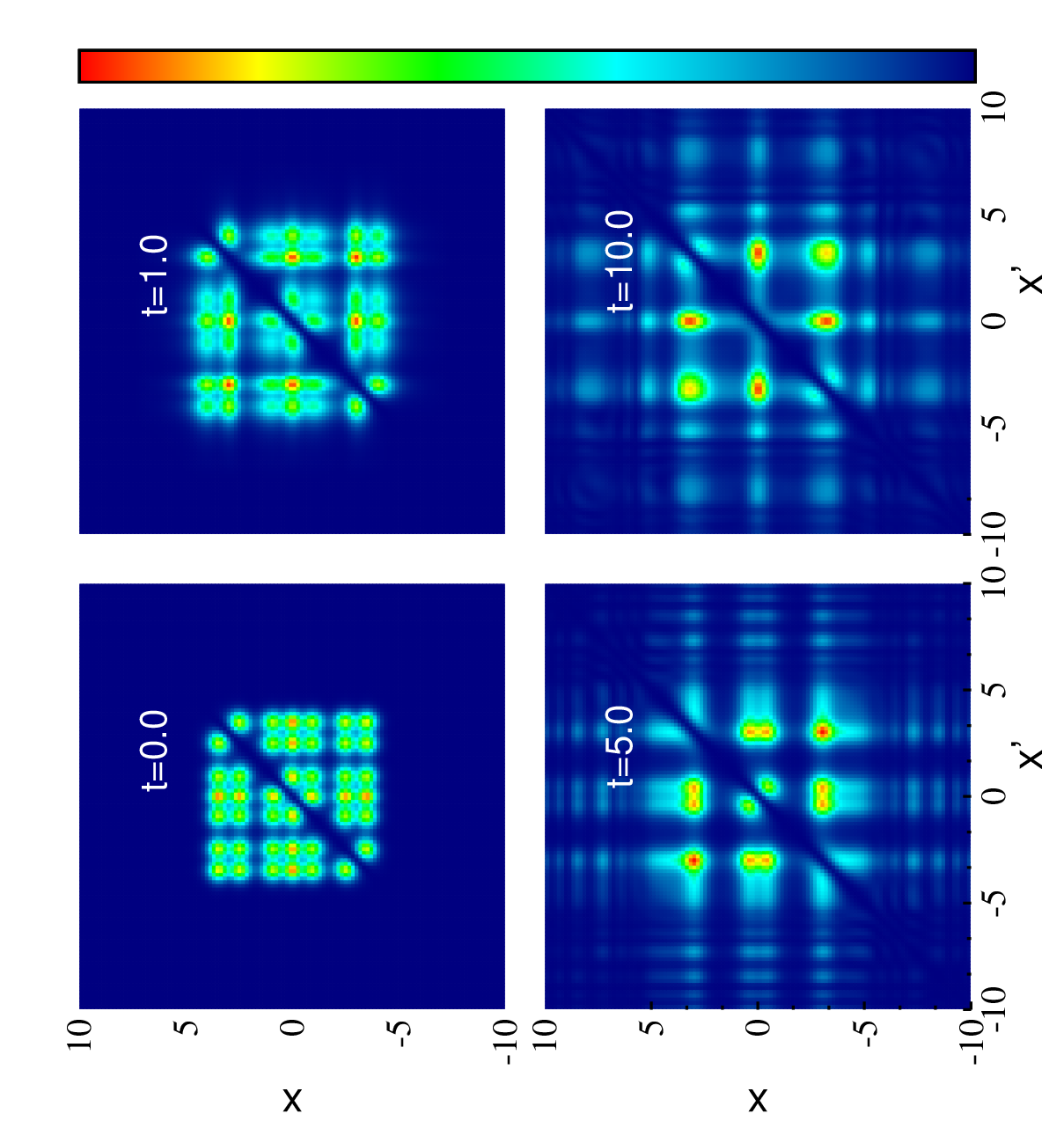}
\caption{Dynamics of the reduced two-body density matrix $\rho^{(2)}(x,x^{\prime})$ post quench for interacting fermions with unit filling factor, $N=7$ fermions expands in $S=7$ lattice sites. The interaction strength is $g_d=1.0$, the stronger dipolar interaction enhances expansion, the many-body features of the core cloud become complete distorted signifying that expansion happens from the central lattice. Computation is done with $M=10$ orbitals.  See the text for discussion.
}
\label{fig9}
\end{figure}

In Fig.~\ref{fig7}-~\ref{fig9}, we present the emergent two-body correlations by the measures of reduced two-body density matrix for unit filling case, $N=7$ dipolar fermions in $S=7$ sites, maintaining the previous quench protocol. We consider three specific cases of interaction strength; weak repulsion $g_d=0.02$ (Fig.~\ref{fig7}), stronger interaction $g_d=0.2$ (Fig.~\ref{fig8}) and very strong interaction $g_d=1.0$ (Fig.~\ref{fig9}). We choose some selected time points $t=0.0, 1.0, 5.0, 10.0$ to manifest the many-body features in the short-time dynamics. The figures represent how the long-range repulsion impacts the correlation spreading across the lattice.

For $g_d=0.02$, we present the correlation dynamics in Fig.~\ref{fig7}. At $t=0.0$, the central bright puddle with a distinct correlation hole signifies three fermions residing in the central lattice; two distinct lobes with clear extinction of diagonal correlation in the adjacent sites explain two fermions sitting in the left well and two are in the right well. In the postquench dynamics, at $t=1.0$, the correlation spreads, correlation hole expands, however the off-diagonal corelation is not affected seriously. At $t=5.0$, when the correlation spreads beyond the three middle sites, low-density cloud appears in the distant lattice sites and the off-diagonal correlation becomes stronger. At $t=10.0$, the correlation structure across the three middle sites remain unchanged, only the low-density cloud reaches the lattice boundary. We conclude that $g_d$ being very small, does not exhibit profound role in the correlation dynamics. The central correlation remains unaffected only some far distant correlation appears indicating transport of small fraction of atoms in the outer lattice sites. 

For higher interaction strength, $g_d=0.2$, we initially observe that corelation is confined within three middle wells as presented in Fig.~\ref{fig8}. In the course of time, we find some trace of cloud in the distant lattice sites at $t=5.0$. The cloud density is already weak compared to the previous case of weak dipolar repulsion. At $t=10.0$, we find distant cloud disappears and the correlation becomes confined strictly in the three middle sites. It indicates that stronger repulsion hinders the correlation spreading. The observation is also in good agreement with the cloud size measurement as presented in Fig.~\ref{fig6}.

The correlation spreading for very strong interaction, $g_d=1.0$, is presented in Fig.~\ref{fig9} for the same time points. Initially, the correlation in the central lattice is distinct compared to the previous cases of weak and strong interaction. The bright lobes around the central lattice also exhibits three bright spots clearly signifying the presence of three fermions. Four other fermions are distributed in the left and right lattice with distinct correlation hole. At $t=1.0$, the cloud expands as a whole and at the same time, correlation in the central lattice is diminished. At $t=5.0$, the expanding cloud reaches at the lattice boundary. At the same time, the off diagonal correlation around the central lattice becomes distorted facilitating the transport of fermions in the distant lattice sites. At much longer time, $t=10.0$, the expansion is strongly enhanced at the cost of initially concentrated correlation in the three middle wells. 

\section{Conclusions}

Ultracold fermions in optical lattices provide an ideal platform to study the fascinating features of nonequilibrium dynamics. They offer unprecedented real-time control of almost all relevant parameters and act as a quantum simulator of strongly correlated many-body systems. The last decade has seen outnumbered investigations both theoretical and experimental to understand the transport properties of the expanding cloud upon quenching the trapping potential in higher dimension and observing crossover from ballistic to bimodal expansion. Less attention has been paid for lower dimension when the effect of quantum corelation plays a significant role.

Little is known about the interplay of long-range interaction in the expansion dynamics of the ultracold dipolar fermions following a quench of trapping potential. Besides the importance of incorporating long-range interaction in solid-state material, the recent access of versatile platforms of atomic and molecular species in the ultracold labs demands the study of quantum transport of interacting dipolar fermions in out-of-equilibrium dynamics. We numerically investigate the expansion of initially localized dipolar fermions in one-dimensional optical lattice. The out-of-equilibrium process is initiated on sudden switching off the harmonic confinement. Our study considers different filling factors, covering the entire range of weak to strong interaction, repulsive as well as attractive. We find stringent effect of long-range dipolar interaction in the expansion of fermionic cloud. In the noninteracting limit, we find ballistic expansion of the fermions, however the corresponding two-body density exhibits how the fermions re-localize themselves during expansion to satisfy Pauli principle. The root mean square radius of the expanding cloud exhibits some aperiodic fluctuations, reaching the lattice boundary and reflects back. For interacting fermions the interplay of dipolar interaction is more complex. We find stringent many-body features in the expansion dynamics. However attractive interaction favors to stabilize cluster states. We also observe that stronger dipolar interaction hinders the spreading correlation and thus exhibiting reduction of expanding cloud size associated with almost periodic breathing dynamics. At variance, very strong interaction enhances expansion and cloud size significantly increases irrespective of the filling factor. 
The observed dynamics can be detected in the expansion with dipolar fermions in 1D geometry, and it would be very interesting to compare 1D results with the findings obtained in higher dimensional setups. 

\section*{Acknowledgments} 
BC acknowledges significant discussion of results with P. Molignini and S. Mistakidis.
BC and AG thank Fundação de Amparo à Pesquisa do Estado de São Paulo (FAPESP), grant nr.~2023/06550-4. AG also thanks the funding from Conselho Nacional de Desenvolvimento Científico e Tecnológico (CNPq), grant nr.~306219/2022-0. 

\appendix
\section{Orbital convergence for interacting fermions}

\begin{figure}
\centering
\includegraphics[width=\linewidth]{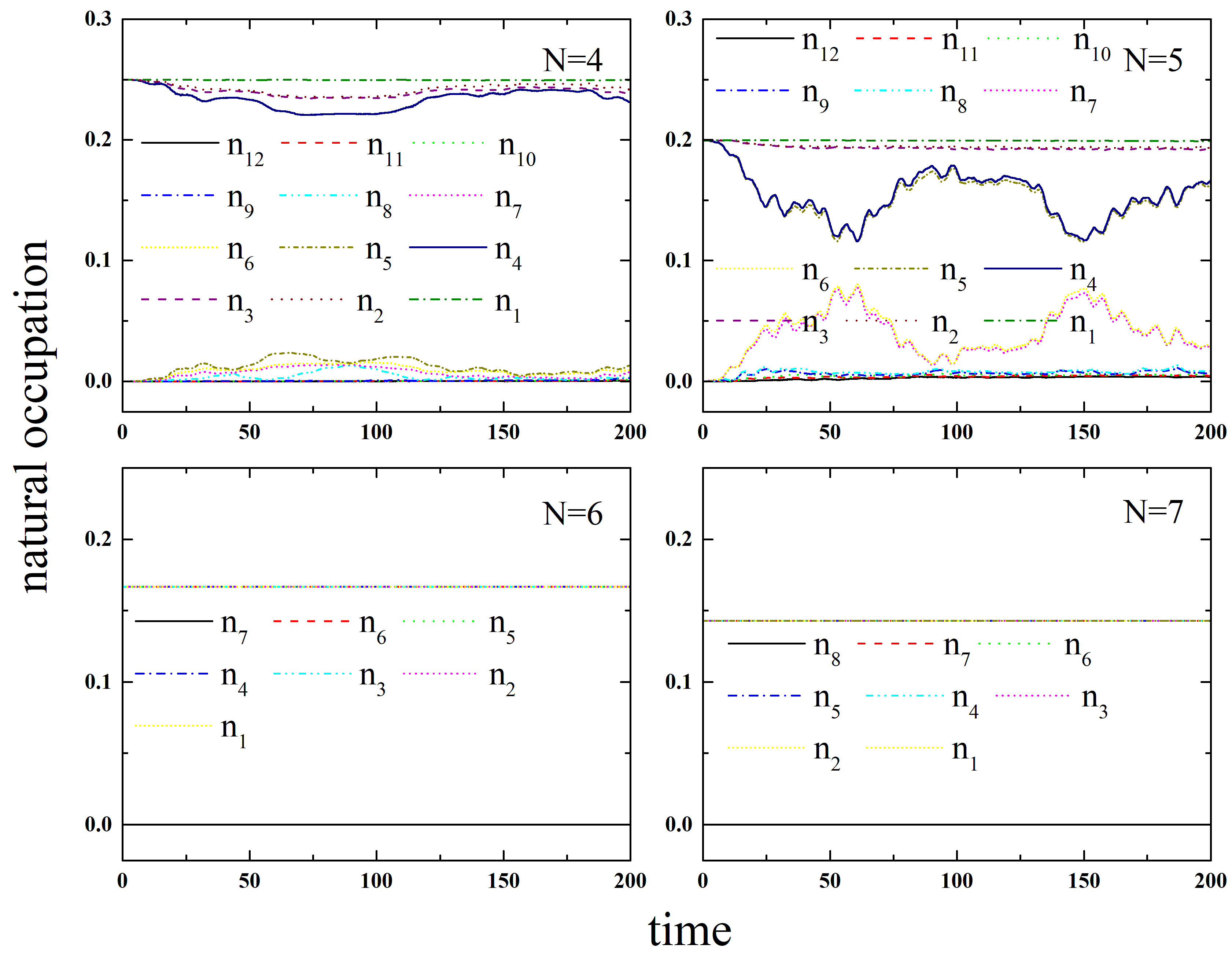}
\caption{ Time evolution of orbital occupation in the post quench dynamics for different number $(N)$ of fermions and interaction strength $g_d=0.2$. For $N=4$ and $N=5$, we utilize $M=12$ orbitals, those are sufficient to achieve convergence. For the system with $N=6$ and $N=7$ fermions, computation is done with $M=10$ orbitals. However for clarity, for $N=6$ fermions, occupation in first seven orbitals and for $N=7$ fermions, occupation in first eight orbitals are presented. It clearly demonstrates that the last orbital population is zero in the entire dynamics. See the text for discussion.
}
\label{fig10}
\end{figure}

\begin{figure}
\centering
\includegraphics[width=\linewidth]{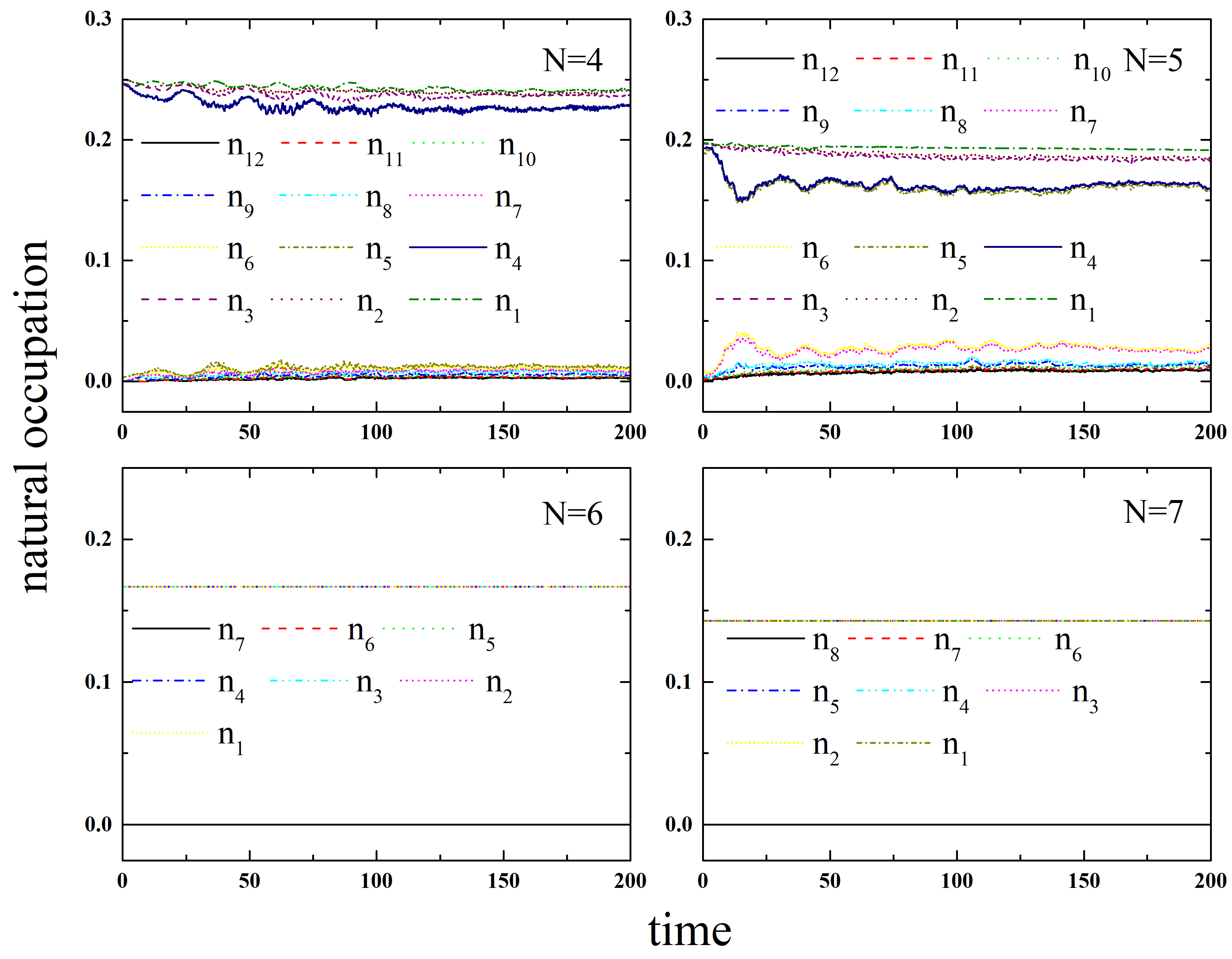}
\caption{ Time evolution of orbital occupation in the post quench dynamics for different number $(N)$ of fermions and higher interaction strength $g_d=1.0$. For $N=4$ and $N=5$, we utilize $M=12$ orbitals, those are sufficient to achieve convergence. For the system with $N=6$ and $N=7$ fermions computation is done with $M=10$ orbitals. However for clarity, for $N=6$ fermions, occupation in first seven orbitals and for $N=7$ fermions, occupation in first eight orbitals are presented. It clearly demonstrates the last orbital population is zero in the entire dynamics. See the text for discussion.
}
\label{fig11}
\end{figure}

The MCTDH-X ansatz becomes numerically exact in the limit of $M \rightarrow \infty$. However very often it is possible to achieve fully converged and thus numerically exact results utilizing finite number of orbitals. It not only guarantees to have ground state properties but also can have converged dynamics. The important criterion is to check the orbital population until the last orbital occupation becomes negligible. 

In the main text, we have reported the time evolution of orbital occupation for the noninteracting cases with various filling factor. $M=N+1$ orbitals were sufficient to describe the converged density dynamics of $N$ noninteracting fermions. In Fig.~\ref{fig10}, we report full time evolution of the orbital occupation for particle numbers $N=4,5,6,7$ for a specific interaction strength $g_d=0.2$. We observe strong dependence of orbital occupation on the filling factor. When filling factor is close to one, $(N=6)$, each orbital exhibits 16.66 \% population. For filling factor one $(N=7)$, each orbital contributes 14.28 \%. Thus computation with $M=7$ orbitals for $N=6$ fermions and $M=8$ orbitals for $N=7$ fermions would be sufficient for converged results. However we carry on simulation with $M=10$ orbitals and present orbital occupation for the first $M=7$ orbitals for $N=6$ and first $M=8$ orbitals for $N=7$. It is clearly manifested that the last orbital population is zero. For $N=4$, there is a clear dominance of first four orbitals. We utilize $M=12$ orbitals to ensure that the last orbital occupation is insignificant. For $N=5$, first three orbitals exhibit significant population, however other higher orbitals also contribute. To compromise computational complexity and orbital convergence, we have utilized $M=12$ orbitals. However, we observe a typical feature of oscillation in the dynamics exhibited by fourth-fifth and sixth-seventh orbitals. 

Following Fig.~\ref{fig10}, we present the time evolution of orbital occupation  with higher interaction strength $g_d=1.0$ in Fig.~\ref{fig11}. As before for particle number $N=6$ and $7$, the number of orbitals used in the computation is $M=10$. However we plot first $M=N+1$ orbitals in each case and convergence is established. For $N=4$ and $5$, we utilize $M=12$ orbitals for fully converged results. For $N=4$, as before, lowest first four orbitals exhibit dominating effect, whereas For $N=5$, apart from first three orbitals, other higher orbitals also contribute. However the typical feature of oscillation as shown in Fig.~\ref{fig10} is not observed.

\section{Density dynamics for interacting fermions with higher filling factor}

\begin{figure}
\centering
\includegraphics[scale=0.2, angle=-90]{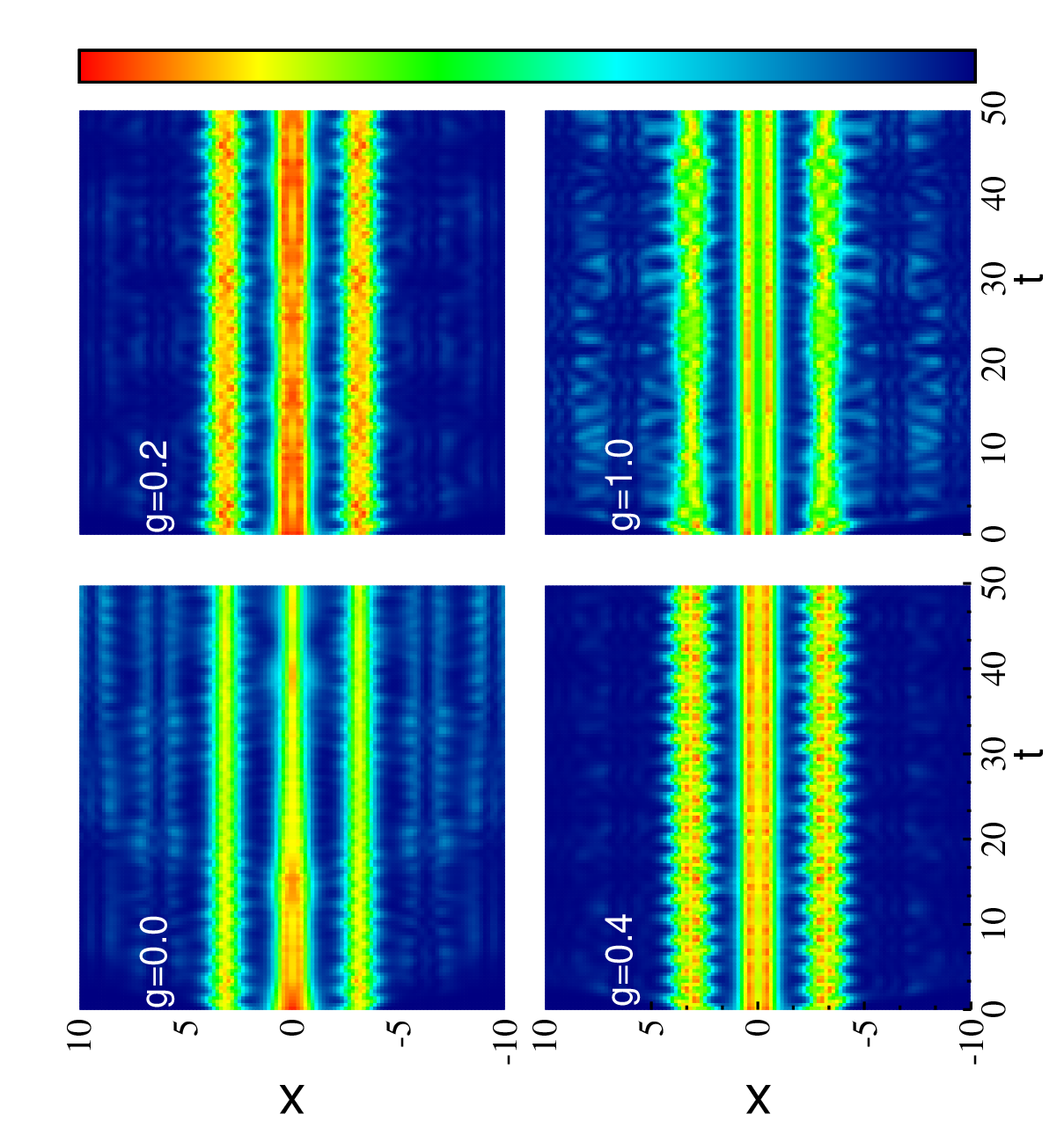}
\caption{ Dynamics of reduced one-body density $\rho(x,t)$ post quench for $N=6$ fermions for four specific choices of $g_d=0.0,0.2,0.4, 1.0$. The noninteracting fermions expand as expected, however for stronger dipolar interaction $g_d=0.2$ and $0.4$, the outer lattice expansion is arrested. The central lattice exhibits intriguing features in the dynamics. For very strong interaction $g_d=1.0$, the long-range repulsion enhances outer lattice expansion and the central lattice presents the independent dynamics of two fermions. Computation is done with $M=10$ orbitals. See the text for discussion.
}
\label{fig12}
\end{figure}

\begin{figure}
\centering
\includegraphics[scale=0.2, angle=-90]{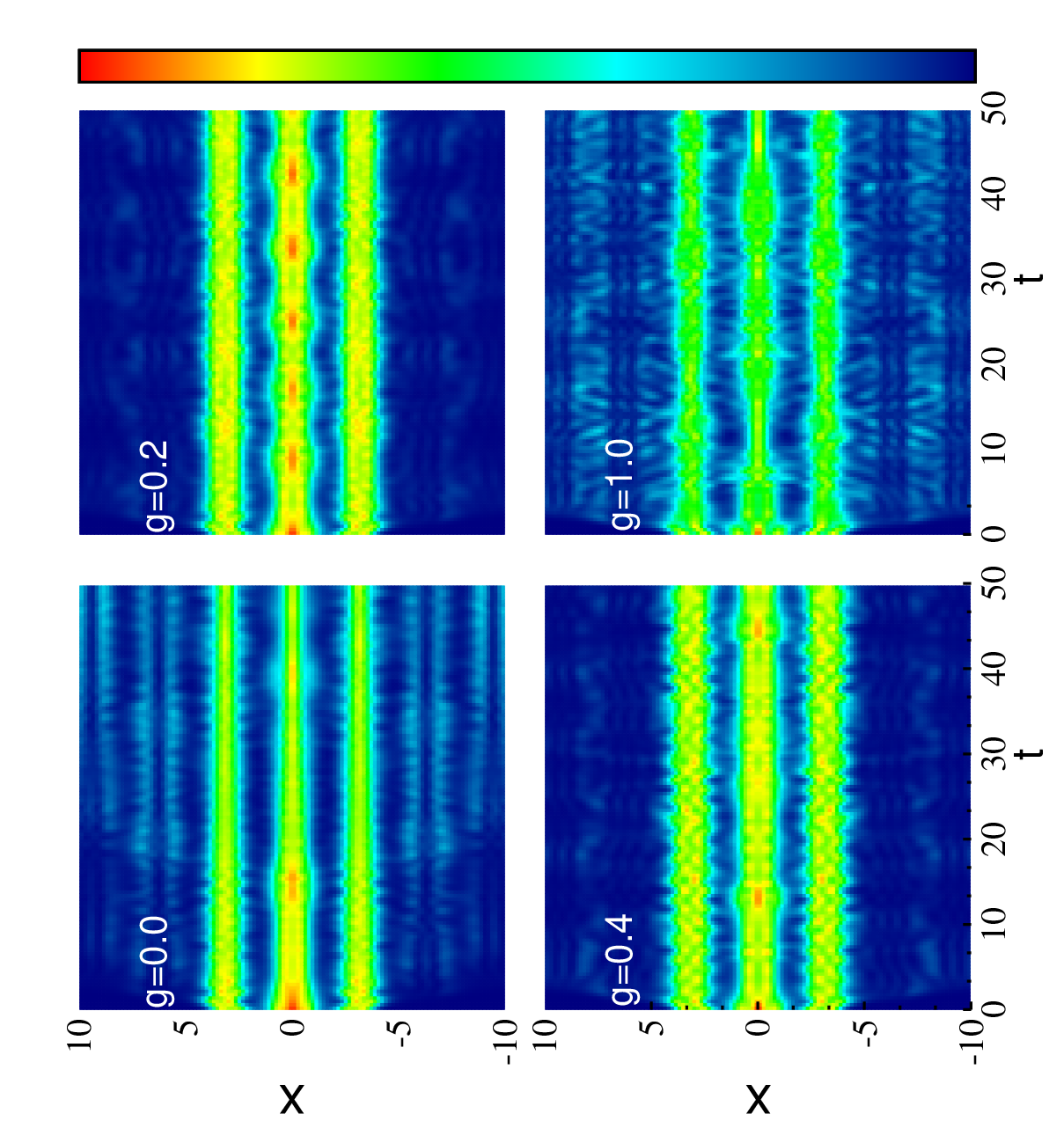}
\caption{ Dynamics of reduced one-body density $\rho(x,t)$ post quench for $N=7$ fermions for four specific choices of $g_d=0.0,0.2,0.4, 1.0$. The noninteracting fermions expand as expected, however for stronger dipolar interaction $g_d=0.2$ and $0.4$, the outer lattice expansion is arrested. The central lattice exhibits intriguing features in the dynamics. For very strong interaction $g_d=1.0$, the long-range repulsion enhances outer lattice expansion, the initial features in the central lattice are also lost signifying that fermions from the inner lattice also participate in the expansion. Computation is done with $M=10$ orbitals. See the text for discussion.
}
\label{fig13}
\end{figure}

In this section, we present the one-body density dynamics $\rho(x,t)$ for interacting fermions with higher filling factors. In the main text, we have discussed expansion dynamics for $N=4$ interacting fermions. We have observed that for weakly interacting fermions, the long-range interaction is sub-relevant, the fermions expand due to their density gradient. For stronger interaction, the long-range repulsion plays a dominating role, hinders the expansion, the density is confined within the three middle wells only. However, stringent many-body features like breathing dynamics and oscillating fermionic cloud is manifested. 

This section is devoted to understand the leading role of long-range interaction when the filling factor is close to one ($N=6$) and exactly one ($N=7$). The density dynamics for $N=6$ fermions is presented in Fig.~\ref{fig12} and for $N=7$ fermions in Fig.~\ref{fig13} for $g_d=0.0, 0.2, 0.4, 1.0$. For $g_d=0.0$, the high-density cloud localized in three middle wells is surrounded by low-density cloud spreading upto lattice boundary. The same dynamics is observed for $N=7$ fermions presented in Fig.~\ref{fig13}. For $N=6$ with $gd=0.2$, the expansion in the lattice boundary is drastically reduced, six interacting fermions are almost equally distributed in the three middle wells and unclear expansion-contraction dynamics is manifested in the central well. Whereas for $N=7$ interacting fermions with $g_d=0.2$, clearly exhibits the breathing dynamics of fermionic cloud in the central lattice. With much higher interaction, $g_d=0.4$, the outer-lattice expansion is more hindered. For $N=6$ fermions, two separated bright jets are visible in the central lattice in the entire dynamics. This signifies that a pair of fermions carry their self identity in each well, whereas very fast breathing oscillation of fermionic cloud is demonstrated in the left and right wells. For $N=7$, atoms are not equally distributed in three wells, thus faded irregular jets are appeared in the density dynamics. For very strong interaction, $g_d=1.0$, long-range repulsion enhances the expansion to outer lattice in contrast to $N=4$ strongly interacting fermions as discussed in Fig.~\ref{fig5}. For $N=6$, the existence of pair of fermions in the central lattice is maintained, however fermions from outer lattice transport to lattice boundary easily due to strong long-range repulsion and some interference pattern is observed. The situation becomes more complex for $N=7$ strongly interacting fermions, which are not uniformly distributed in the three wells. Thus the fermions from central as well as middle lattice sites expand immediately after removal of the confinement. The features in the central lattice is lost due to the fermionic transport.

\section{Radius of expanding cloud in varying lattice size}

\begin{figure}
\centering
\includegraphics[width=\linewidth]{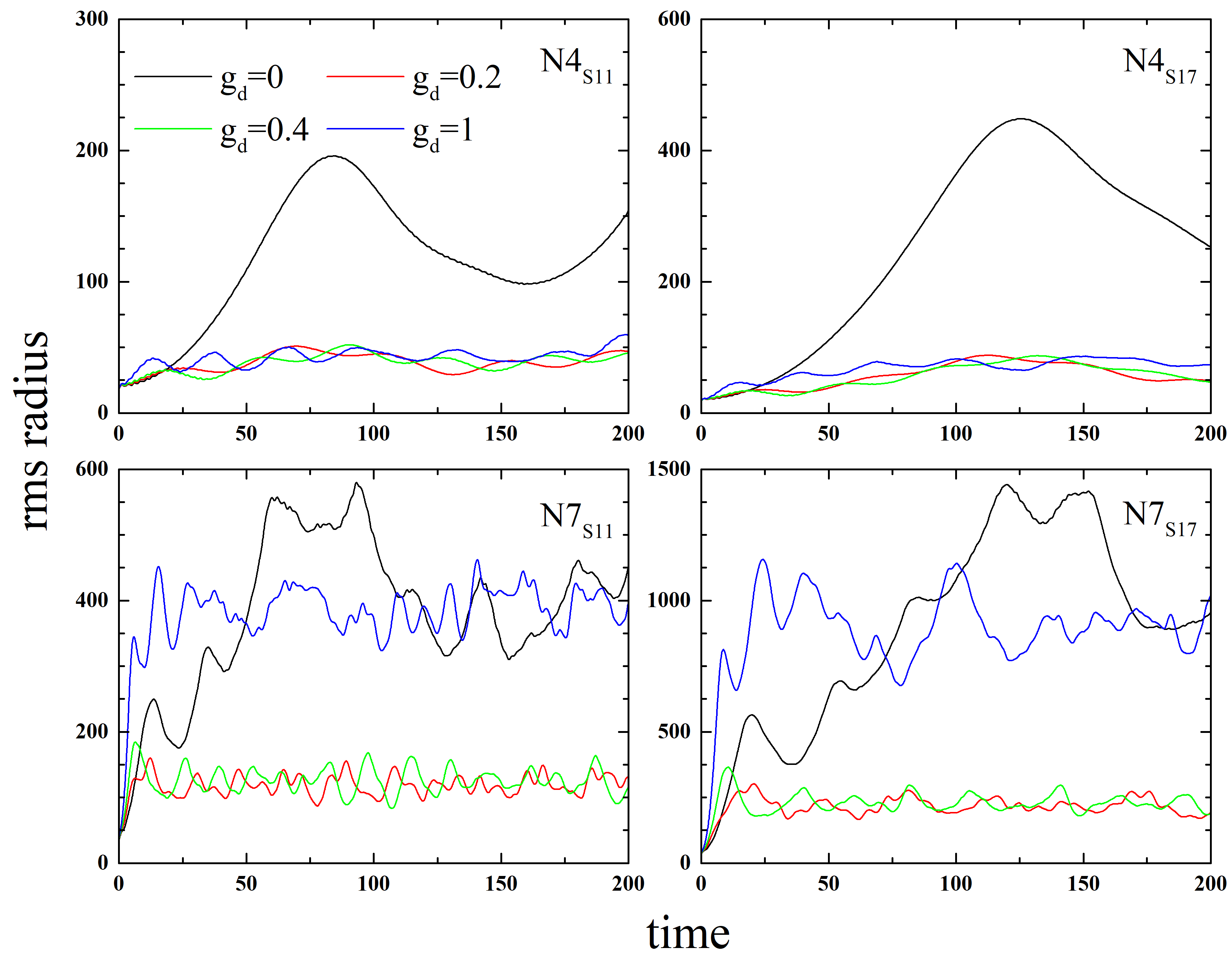}
\caption{ Time evolution of root mean square radius of the fermionic cloud for different lattice size and for four specific choices of interaction strength $g_d=0.0, 0.2,0.4,1.0$. Top panel represent size of the expanding cloud of $N=4$ fermions in $S=11$ and $17$ sites, conclude the same physics for $S=7$ sites as discussed in the main text. Bottom panel presents the same for $N=7$ fermions in $S=11$ and $17$ sites and observation supports the conclusion made in the main text for smaller lattice size. For both system $M=12$ orbitals are used. See the text for discussion.
}
\label{fig14}
\end{figure}

In this Appendix, we present Fig.~\ref{fig6} results of the main text for larger lattice size to demonstrate that the main text can be generalized. Fig.~\ref{fig6} has concluded that size of the expanding cloud strongly depends on the filling factor. For all cases, the noninteracting fermions and weakly interacting dipolar fermions expand due its diluteness and cloud size increases with increase in filling factor. However, stronger dipolar interaction hinders the expansion and very strong repulsion enhances the expansion. Although, for the low filling factor ($N=4)$, the enhancement in cloud size for strong dipolar interaction has not been observed. Whereas, for $N=5,6,7$, Fig.~\ref{fig6} has demonstrated the unique feature how the long-range interaction initially prohibits the expansion whereas stronger interaction facilitates the expansion.

The present appendix is aimed to demonstrate the same physics for larger lattice size. We consider two specific cases with atom number $N=4$ and $7$ and two specific cases of lattice size,  $S=11$ and $S=17$. The initial and final protocols remain same as in the main text. For all cases of noninteracting fermions, the cloud expands as expected, however larger lattice provides larger space, thus the cloud size is maximum in $S=17$ sites. For $N=4$, the broad peak is observed as in Fig.~\ref{fig6}, signifying that expanding cloud hit the lattice boundary, cloud size becomes maximum and then reflects back. The cloud size is more fluctuating for $N=7$ noninteracting fermions indicating very fast reflection from the lattice boundary and it is aperiodic. We have observed the same physics in the main text with $S=7$ lattice sites. For higher interaction strengths, the expansion is hindered due to long range repulsion. For $N=4$ in $S=11$ sites, with $g_d=0.2$ and $0.4$ the average cloud size remains $\simeq 25$ as observed for $S=7$ lattice sites. Even for very strong dipolar interaction, $g_d=1.0$, the cloud size remains same as observed in Fig.~\ref{fig6}. For $N=7$, with $g_d=0.2$ and $0.4$, the cloud expansion is restricted as observed in smaller lattice size, the average cloud size varies between $100$ to $110$, which is comparable to the cloud size for $S=7$ sites. For much stronger interaction, the long-range repulsion enhances spreading and the cloud size increases significantly. The same physics is observed for larger lattice size $S=17$.

\bibliography{ref}

\end{document}